\documentclass{elsarticle}
\usepackage{graphicx,color}
\usepackage{amsmath}
\usepackage{amssymb}

\def\nab{{\boldsymbol \nabla}}

\def\simle{\mathrel{\rlap{\raise 0.511ex \hbox{$<$}}{\lower 0.511ex \hbox{$\sim$}}}}
\def\simge{\mathrel{ \rlap{\raise 0.511ex \hbox{$>$}}{\lower 0.511ex \hbox{$\sim$}}}}

\newcommand{\rmd}{{\rm d}}

\newcommand{\nn}{\nonumber\\ }

\newcommand \beq{\begin{eqnarray}}
\newcommand \eeq{\end{eqnarray}} 
\def\simle{\mathrel{\rlap{\raise 0.511ex \hbox{$<$}}{\lower 0.511ex 
\hbox{$\sim$}}}}
\def\simge{\mathrel{ \rlap{\raise 0.511ex 
\hbox{$>$}}{\lower 0.511ex \hbox{$\sim$}}}}

\def\p{{\boldsymbol p}}

\begin{document}
\begin{frontmatter}
\title{Gluon Transport Equations with Condensate \\ in the Small Angle Approximation}
\author[cea]{Jean-Paul Blaizot}
\author[ind,rbrc]{Jinfeng Liao}

\address[cea]{Institut de Physique Th\'{e}orique (IPhT),  CNRS/URA2306, CEA Saclay,\\
F-91191 Gif-sur-Yvette, France}
\address[ind]{Physics Department and Center for Exploration of Energy and Matter,
Indiana University, 2401 N Milo B. Sampson Lane, Bloomington, IN 47408, USA}
\address[rbrc]{RIKEN BNL Research Center, Bldg. 510A, Brookhaven National Laboratory,
   Upton, NY 11973, USA}

\begin{abstract}
We derive the set of kinetic equations that control the evolution   of gluons in the presence of a condensate. We show that the dominant singularities remain logarithmic when the scattering involves particles in the condensate. This allows us to define a consistent small angle approximation. 
\end{abstract}
\end{frontmatter}

\section{Introduction}

In previous papers\cite{Blaizot:2011xf,Blaizot:2013lga}, the Boltzmann equation including proper treatment of Bose statistics, also referred to as the Boltzman-Nordheim equation, was used to describe the evolution of a dense system of gluons, such as that produced in the early stages of an ultra-relativistic heavy ion collision.  The computations were done by keeping only elastic scattering in the collision kernel. Also, the small scattering angle approximation was used, thereby reducing the Boltzmann equation to a Fokker-Planck equation, much easier to solve.  A surprising conclusion of this analysis was that for a wide variety of initial conditions appropriate for ultra relativistic heavy ion collisions, Bose condensation occurs. This is because the density of gluons is too large, with respect to their total energy, to be accommodated by a Bose distribution in equilibrium.  One might object that higher order inelastic scattering terms might alter this conclusion, but computations to leading order in inelastic processes suggest that, at least at early times, particle number is approximately conserved, and Bose condensation is favored\cite{Huang:2013lia} (see also discussion in \cite{BJL} and further references therein).

 The analysis to date is incomplete since terms which explicitly depend upon the condensate have been ignored
 in the scattering term of the transport equation (aside from the inelastic processes just mentioned).  Including such contributions is essential to describe the evolution of the system beyond the onset of condensation, i.e., with a condensate present.  These contributions allow for transport of particles into and out of the condensate. It is the purpose of this paper to derive the form of the transport equation in the presence of the condensate, in the small scattering angle approximation. In order to do so, we  need to include the effects of gluon mass and Debye screening in order to regulate the infrared and obtain unambiguous results\footnote{The authors of Ref.~\cite{Xu:2014ega} managed, by a suitable choice of the collision kernel, to give a meaning to the massless limit.}. The relevant modifications of the collision kernel that are due to finite gluon masses are discussed in a companion paper \cite{BJL}. The equations that describe the evolution in presence of a condensate have already been obtained in other contexts \cite{Semikoz:1994zp,Lacaze:2001qf,Spohn}. The present  paper analyzes the singularities that originate from the scattering of zero momentum particles when the interactions are long range, and therefore small angles scattering dominate. We show that linear singularities that emerge in intermediate stages of the calculation eventually cancel, leaving us with logarithmic singularities only. This allows us to obtain a consistent small scattering angle approximation, also referred to as ``leading log'' approximation. 
 
The general structure of the transport equation in the presence of a condensate is reviewed in the next section, where the final set of leading log equations, which constitute the main result of this paper, are given. In Sect.~\ref{Sect:Ci} we present the calculation of the individual contributions to the collision kernel, keeping in the computation the leading and sub-leading singularities. The cancellation of the linear singularities is established in Sect.~\ref{Sect:leading log}, where the consistency of the resulting leading log equations is verified by analyzing conservation laws of particle number and energy. 

\section{The collision term in the presence of  a condensate}

We start with the transport equation including only $2\to 2$ elastic scattering, i.e. 
\begin{eqnarray} \label{Eq_transport}
 \partial_t f_1&  = &  \frac{1}{2} \int {{\rmd^3\p_2} \over {(2\pi)^3 2 E_2}}{{\rmd^3\p_3} \over {(2\pi)^3 2 E_3}}
   {{\rmd^3\p_4} \over {(2\pi)^3 2 E_4}} {1 \over {2E_1}}    \nonumber \\
   & \times & \mid {\cal M}_{12\to 34} \mid^2(2\pi)^4 \delta^{(4)}(P_1 + P_2 - P_3 -P_4)  \nonumber \\
   &\times & \{f_3f_4(1+f_1)(1+f_2) -  f_1f_2(1+f_3)(1+f_4)  \}\nonumber\\
  \end{eqnarray}
    where the factor $1/2$ in front of the integral is a symmetry factor, and $f_i$ is a shorthand for $f_{\p_i}=f(\p_i)$, and similarly, $E_i=E_{\p_i}$. Summation over color and polarization is performed on the gluons 2,3,4; average over color and polarization is performed for gluon 1. We use the small scattering angle approximation for the matrix element for gluon-gluon scattering ($1+2\to 3+4$),  with the dominant contributions arising  from the $t$ and $u$ exchange channels:  
 \begin{eqnarray}
 \mid {\cal M} \mid^2  \approx  -\kappa \left[   \frac{s\, u}{(t- m_D^2)^2} + \frac{t\, s}{(u- m_D^2)^2} \right] \, , \quad \kappa\equiv 128\pi^2\alpha_s^2 N_c^2  \,\, ,
 \end{eqnarray}
 where $s,t,u$ are the standard Mandelstam variables:
 \beq s=(P_1+P_2)^2,\qquad t=(P_1-P_3)^2,\qquad u=(P_1-P_4)^2,
 \eeq
 and $P_i$ denotes the four-momentum of particle $i$, $P_i=(E_{\p_i},\p_i)$. Momentum and energy conservations imply $P_1+P_2=P_3+P_4$.  
We have introduced a screening mass $m_D$ to regulate the infrared behavior of the one gluon exchange. Note that this screening mass is different from the external gluon mass $m$ entering the dispersion relation of the colliding gluons, $E_\p=\sqrt{\p^2 + m^2}$. Both masses are supposed to be small compared to the typical momenta of the particles that are taking part in the collision. 

Our main goal in this paper is to obtain the form of the kinetic equations that govern the evolution of the distribution function when a condensate is already formed and participates in 
the collisions. In this case the distribution function is generically of the form 
\begin{eqnarray}\label{distrwithcond}
f(\p) = n_c \, (2\pi)^3 \delta^{(3)}(\p)+ g(\p)
\end{eqnarray}
 with $g(\p)$ referred to as  the ``regular'' part of the distribution. Note however that as soon as a condensate is present, $g(\p)$ is expected to be singular at $p=0$. In equilibrium,  $g(\p)$ would be a Bose distribution with chemical potential equal to the mass, i.e., $g(p)\simeq 2mT/p^2$ at small $p$. We have verified in a previous work that this singular behavior is indeed approached as the system gets close to the onset of condensation. We shall assume here  that, when the condensate is formed, the distribution $g(p)$ keeps a singular component in $1/p^2$. 
 
 By inserting this two-component distribution back into the transport equation, we get
\begin{eqnarray}\label{distrwithcondeq}
(2\pi)^3 \delta^{(3)}(\p_1) \, {\partial}_t n_c +   {\partial}_t g(\p_1) =C[n_c,g].  
\end{eqnarray}
Let us examine the collision integral $C[n_c,g]$ on the right hand side of this equation.  According to Eq.~(\ref{distrwithcond}),  each distribution $f_i$ in the collision integral gives two contributions, one from the condensate, which comes with the delta function $\delta^{(3)}(\p_i)$,  and one associated with the  regular function $g(\p_i)$.
\begin{enumerate}
\item
 If all four distributions $f_{1,2,3,4}$ pick up the $\delta^{(3)}(\p_i)$ component, then they become identical and the gain and loss terms cancel each other. 
 
 \item The same situation prevails if three distributions pick up the $\delta^{(3)}(\p_i)$ component, since then the fourth must also be at zero momentum due to the overall energy momentum conservation. 
 
\item If two distributions pick up the $\delta^{(3)}(\p_i)$ component, then there are two situations: (a) one of the factor $\delta^{(3)}(\p_i)$ is in the initial state and the other one is in the final state ---say $\p_1=\p_3=0$ --- in this case energy momentum conservation imposes the other two to have equal momenta, which results in the  cancellation between the gain and loss terms; (b) both factors $\delta^{(3)}(\p_i)$ are in either initial or final states ---say $\p_1=\p_2=0$ --- in this case the other two momenta have to vanish too in order to satisfy the energy momentum conservation, and we are back to the situation where all four momenta vanish, for which the collision term also vanishes.

\end{enumerate}
We therefore need to deal only with the terms that contain one or no delta-function component. We can write the collision integral as a sum over five terms: 
\begin{eqnarray}
C[n_c,g] = C_0[g] + \sum_{i=1}^4  C_i[n_c, g] . \label{defC}
\end{eqnarray}
The term $C_0$ has no condensate contribution and the remaining four  $C_i$'s have a condensate contribution, with $p_i=0$ in $C_i$.

The next step involves projecting the full transport equation into a set of two coupled evolution equations for the condensate and for the regular distribution. There is a subtlety here, since, as we shall see, contributions to the condensate density can arise from several places.  To handle this properly, we  define the condensate density $n_c$ by integrating the distribution function is a small sphere of radius $p_0$ and let $p_0$ go to zero. That is, 
\begin{eqnarray}\label{ncdefinition}
n_c = \int_0^{\p_0\to 0} p^2dp \int \frac{d\Omega_{\hat \p}}{(2\pi)^3} f(\p).
\end{eqnarray} 
 We use this procedure to obtain the equation for the condensate density 
\begin{eqnarray} \label{eq_dt_nc}
{\partial}_t n_c   = \tilde{C_1}[n_c,g] + \int_0^{\p_0\to 0} p^2dp \int \frac{d\Omega_{\hat \p}}{(2\pi)^3} \left[   C_0+ C_2+C_3+C_4\right],
\end{eqnarray}
with $C_1[n_c, g] =  (2\pi)^3 \delta^{(3)}(\p_1) \,  \tilde{C_1}[n_c,g]$. 
While $C_1$ represents an obvious contribution since the delta function is explicit, we shall  see that some of the contributions $C_{0,2,3,4}$ are singular as $\p_1 \to 0$ and give also non vanishing contributions to the right hand side of this equation.  
For $|\p_1|>0$, the transport equation reads 
\begin{eqnarray} \label{eq_dt_g}
 {\partial}_t g(\p_1) = C_0[g] +   C_2[n_c, g] + C_3[n_c, g] + C_4[n_c, g].
\end{eqnarray}
In what follows, we will derive these terms   one by one. \\

But before we proceed with the calculation of these individual contributions, it is useful to  present the final result. In the presence of the condensate, the kinetic equation gives rise to two coupled equations, which can be written as a single equation of the following form (for an isotropic system, with $p_1=|\p_1|$)
\beq\label{transportfinal}
 \frac{\partial f(p_1)}{\partial t} = -\nab\cdot{\cal J}_0(p_1)  - \frac{n_c}{m}\,  \nab\cdot{\cal J}_c(p_1) \label{eq_log_g0} ,
\eeq
where the current ${\cal J}_0$ is a current that depends only on the regular parts of the distribution functions and whose expression is identical to what it would be in the absence of the condensate (see Eq.~(\ref{currentJ0}) below), while ${\cal J}_c$ which multiplies $n_c$ in Eq.~(\ref{transportfinal}) is given by  
\begin{eqnarray}
{\cal J}_c \equiv   - \,  \hat{\p}_1 \, L_c \,  \left[T^* g_1' + g_1(1+g_1) \right],   \qquad  g_1'\equiv \frac{\rmd g_1}{\rmd E_1}=\frac{E_1}{\p_1}\frac{\rmd g_1}{\rmd p_1},
\end{eqnarray}  
and 
\beq\label{defLcmr}
L_c\equiv \frac{2  \kappa }{32\pi  } \,  \ln\left(\frac{\Lambda}{m_r}\right) ,\qquad m_r\equiv \frac{m_D^2}{2 m}.
\eeq 
Finally $T^*=I_a(p_1)/I_b(p_1)$ is an effective temperature given by  the ratio of two integrals of the function $g(p)$ (see Eq.~(\ref{IaIbdef}) below). Note that $T^*$ is independent of $p_1$.
The two equations (\ref{eq_log_g}) and (\ref{eq_log_c}) can be recovered by separating $f(p_1)$ into a delta function and a regular contribution, with the procedure indicated in Eq.~(\ref{eq_dt_nc}). 
These equations  are obtained in the leading order of the small scattering angle approximation, which consists in isolating the leading singularity of the collision kernel as the screening mass tends to zero. As we shall see, there is a delicate interplay in the derivation between  singularities that arise in the matrix element and those that arise form the  behavior at small $p$ of the function $g(p)$.

\section{Determination of the various contributions $C_i$}\label{Sect:Ci}

In this section, we derive the expressions of the various contributions $C_i$ to the collision integral, and we discuss the elimination of the dominant infrared divergences. We consider a system that is uniform and isotropic in momentum space, so that the distribution function $f(\p)$ depends only on the magnitude $p=|\p|$ of the momentum. 

Before we proceed, we note that, except for the contribution $C_0$ which does not depend explicitly on the presence of the  condensate, for all other contributions the presence of the condensate has an impact on the form of the function $g(p)$ at small momentum. Our analysis relies on the assumption, supported by numerical calculations,  that, once the condensate appears,  $g(p)$  does not deviate much from an equilibrium distribution, i.e.,  $g(p) =  1/(e^{(E-m)/T^*}-1) + \delta g$. This function  has a singular behavior $\sim 1/\epsilon$ with $\epsilon=E-m$ when $p\to 0$. We shall also assume that $\delta g(p)$ is regular as $p\to 0$, so that for small $\epsilon$, $g(p)$ takes the form $g(p)\simeq T^*/\epsilon+V^*$, with $T^*$ and $V^*$ two constants.

\subsection{The contribution $C_0$}

The  term $C_0$ corresponds to the usual collision term involving no particle in the condensate. It is independent of $n_c$ and is obtained by replacing all the $f_i$'s in the collision integral by their corresponding regular parts $g_i$: 
\begin{eqnarray}
C_0[g] &=&  \int_{\p_2}\int_{\p_3}\int_{\p_4}   \frac{ (2\pi)^4 \delta^{(4)}(P_1 + P_2 - P_3 -P_4)}{16 E_1 E_2 E_3 E_4}  \mid {\cal M} \mid^2 
 \nonumber \\ 
&& \qquad \times  \{g_3g_4(1+g_1)(1+g_2) -  g_1g_2(1+g_3)(1+g_4)  \} ,
\end{eqnarray} 
where we use the  shorthand to denote momentum integration $\int_{\p_1}=\int \rmd^3\p/(2\pi)^3$. Note that particle 1 is the particle that one follows in the kinetic equation (i.e. we calculate the collision term corresponding to $\partial_t f_1$). 

The small scattering angle approximation of the above expression has been obtained in a companion paper \cite{BJL}. The result takes the form 
\begin{eqnarray}
C_0[g]=-\nab\cdot {\cal J}_0(p_1) ,
\end{eqnarray} 
where the explicit expression of the current ${\cal J}_0$ is given in \cite{BJL}. In the very small momentum regime $p_1 \ll m$, this expression simplifies and reads
\begin{eqnarray}\label{currentJ0}
{{\cal J}_0}|_{p_1\to 0} &\to&  - \kappa_0    \,  \left [ I_a(p_1) g'_1 + I_b(p_1) g_1(1+g_1) \right] \label{eq_S0_p0},\qquad \kappa_0=36\pi \alpha_s^2{\cal L},
\end{eqnarray}
where $\kappa_0$ is a constant factor and the two integrals $I_a(p_1)$ and $I_b(p_1)$  are given by
\begin{eqnarray}\label{IaIbdef}
I_a(p_1) &\equiv& \frac{p_1}{m} \int_\p \frac{E}{p} g (1+g), \\
I_b(p_1) &\equiv& \frac{p_1}{m} \int_\p  \frac{E}{p} \left( - g'(E_p) \right) ,
\end{eqnarray}
where  $g'(E)\equiv d g(E) / d E = (E/p) \partial_p g$.

\subsection{The contribution $C_1$}

This contribution corresponds to the case where the particle 1 is in the condensate ($\p_1=0$), and can be kicked out of it by collisions with particle 2. 
It is obtained by  replacing,  in Eq.~(\ref{Eq_transport}), $f_1$ and $1+f_1$  by $(2\pi)^3 \delta^{(3)}(\p_1) $. 
We get 
\begin{eqnarray}\label{C1tilde}
\tilde{C}_1[n_c,g] &=&2 \frac{  \kappa n_c}{16 m} \int_{\p_2} \int_{\p_3}   \frac{ 2\pi \delta(m + E_2 - E_3 -E_4)}{  E_2 E_3 E_4} \, \frac{(E_2-E_3)(E_2+m)}{[(E_3-m) + m_r]^2} \nonumber \\ 
&& \qquad \times
 \{ g_3 g_4 - g_2 g_3 -g_2 g_4 - g_2 \} ,
 \end{eqnarray}
 where $m_r \equiv m_D^2 / (2m)$ (see Eq.~(\ref{defLcmr})).
  Here $\p_1=0$, $E_1=m$, so that $s=2m(E_2+m)$, $t=2m(m-E_3)$, and $u=2m(E_3-E_2)$. The factor $2$ in front accounts for the equal contributions from the $t$ and $u$ channels (the collision integral is invariant under the exchange of $\p_3$ and $\p_4$ which corresponds to the exchange of the $u$ and $t$ channels). In Eq.~(\ref{C1tilde}), the integration over  $\p_4$ has been done by using the momentum delta function. At this point, we use the remaining energy conservation delta function to carry out the  angular part of the $\p_3$-integral.  The angular part of $\p_2$ can then be integrated trivially. We get (with $\rmd E_i = p_i \rmd p_i / E_i$, and $p_i^2=E_i^2-m^2$)
 \begin{eqnarray}\label{C1tildeb}
\tilde{C}_1[n_c,g]  &=& \frac{\kappa}{16\pi}\frac{n_c}{m}   \int_{\p_2} \int p_3^2 dp_3   \frac{ \Theta(E_2-E_3)}{  (p_2 E_2) (p_3 E_3) }  \frac{(E_2-E_3)(E_2+m)}{[(E_3-m) + m_r]^2}  \nonumber \\ 
&& \qquad \times
 \{ g_3 g_4 - g_2 g_3 -g_2 g_4 - g_2 \} \nonumber \\  
  &=&     \frac{\kappa}{32\pi^3}\frac{n_c}{m}   \int_m dE_2 \int_m^{E_2} dE_3  \,  \frac{p_2^2 - (E_2+m)(E_3-m)}{[(E_3-m) + m_r]^2} \nonumber  \\
  && \qquad  \qquad 
 {\bigg \{ } g_3 g_4 - g_2 g_3 -g_2 g_4 - g_2 {\bigg \} }{\bigg |}_{E_4 \to m+E_2-E_3}.
\end{eqnarray} 
It is not difficult to see that the Bose-Einstein thermal distribution with $\mu = m$ is still a fixed point, i.e., the combination of statistical factors vanishes for $g$ a Bose distribution.  The same remark will apply to all the cases to be discussed in this section, and will not be repeated.

In order to obtain  the small scattering angle approximation of the expression above, we focus on the singular contributions that come from the region  $E_3\gtrsim  m$. This is the region from which we expect a singularity in Eq.~(\ref{C1tildeb}) when $m_D\to 0$. The small scattering angle approximation consists in isolating such singularities, and retaining the leading one as the main contribution to the collision kernel. As we shall see the leading singularity is a logarithmic one. Other singularities emerge, in particular linear singularities, but  we shall show later that these cancel, leaving us with logarithms only. 

To proceed,  we set $\epsilon=E_3-m$ and use the notation $g_\epsilon \equiv g(E_3)$. Then we have  $E_4=m+E_2-E_3=E_2 - \epsilon$, and $g(E_4)\approx g_2  - g'_2 \epsilon + \frac{1}{2} g''_2 \epsilon^2$, where, as before, the prime denotes a derivative with respect to the energy. The difficulty in the analysis comes from the fact that a small angle scattering involves a small momentum transfer, and as a result of such a scattering $p_3\simeq p_1=0$. The singularity of the matrix element as $m_D\to 0$, and the presence of a particle in the condensate, render therefore the collision integral sensitive to the singular behavior at small $p$  of one of the ``regular'' distribution functions, here $g(p_3)$. This is the main subtlety in the analysis.  Using the notation that we have just introduced, we can rewrite Eq.~(\ref{C1tildeb}) in the following form:  
\begin{eqnarray}
\tilde{C}_1[n_c,g]  &=& \frac{\kappa}{32\pi^3}\frac{n_c}{m} \int_m dE_2 \int_0^\Lambda d\epsilon \, \frac{p_2^2 - (E_2+m)\epsilon}{(\epsilon + m_r)^2}  \, \nonumber \\
&& \qquad  \left\{  	\left[  - (\epsilon g_\epsilon) g'_2  - g_2 (1+g_2) \right] +	\epsilon \, \left[ (\epsilon g_\epsilon)\frac{g''_2}{2}  +g_2 g'_2 \right]  \right\},
\end{eqnarray} 
where $\Lambda$ is an ultraviolet cutoff that controls the divergences caused by the  small $\epsilon$ expansion of the distributions (before this expansion, the integrand is naturally cut off at energies $E\sim T$).
By organizing the terms according to the various types of $\epsilon$-integrations, we get
\begin{eqnarray}
\tilde{C}_1[n_c,g]  = \frac{\kappa}{32\pi^3}\frac{n_c}{m} \left\{ -H_a \,L_{g1} - H_b\, L_0+  H_c  \, L_{g2}+ H_d\, L_1  \right\} ,
\end{eqnarray}
with the following finite integrals
\begin{eqnarray}
H_a  & \equiv &  \int p^2 dE_p \, g'(E_p)  = \int p^2 dp  \partial_p g(p)  \\
H_b  & \equiv &  \int p^2 dE_p \, g(E_p)[1+g(E_p) ]  \\
H_c & \equiv &  \int p^2 dE_p \,  \left[ \frac{g''(E_p)}{2} + \frac{(E_p+M)}{p^2} g'(E_p) \right] \\
H_d & \equiv &  \int p^2 dE_p \,  \left[ g(E_p)g'(E_p) + \frac{(E_p+M)}{p^2} g(E_p)\left (1+g(E_p) \right) \right] 
\end{eqnarray}
and the following singular integrals
\begin{eqnarray}
&&L_{g1}\equiv \int_{0}^\Lambda\rmd\epsilon\,\frac{\epsilon g_\epsilon}{(\epsilon + m_r)^2} \approx  \frac{T^*}{m_r} + V^* \, \ln\left(\frac{\Lambda}{m_r}\right)   \\
 &&L_{0}\equiv\int_{0}^\Lambda\rmd\epsilon\, \frac{1 }{(\epsilon + m_r)^2} \approx\frac{1}{m_r}   \\
&&L_{g2}\equiv \int_{0}^\Lambda \rmd\epsilon\, \frac{\epsilon^2 g_\epsilon}{(\epsilon + m_r)^2} \approx T^*\,  \ln\left(\frac{\Lambda}{m_r}\right)   \\
&&L_{0}\equiv\int_{0}^\Lambda\rmd\epsilon\,  \frac{\epsilon }{(\epsilon + m_r)^2} \approx  \ln\left(\frac{\Lambda}{m_r}\right),  
\end{eqnarray} 
where we have kept only the dominant terms when $m_D/\Lambda\to 0$.
To calculate these integrals, we have  adopted the general expansion (see the discussion at the beginning of this section): $g_\epsilon \to T^*/\epsilon + V^* + \hat{o}(\epsilon)$ and therefore  $\epsilon g_\epsilon \to  T^* + V^* \epsilon +  \hat{o}(\epsilon^2)$. Note that terms of order $\hat{o}(\epsilon^2)$ or higher are irrelevant.  

The final result for $\tilde{C}_1[n_c,g]$ can then be written as
\begin{eqnarray}
\tilde{C}_1[n_c,g]  &=& \frac{\kappa}{32\pi^3}\frac{n_c}{m}  \bigg \{  \left( -H_a T^* - H_b \right ) \left( \frac{1}{m_r}\right) \nonumber \\
&& \quad +  
\left( -H_a V^* + H_c T^* + H_d \right )   \ln\left(\frac{\Lambda}{m_r}\right)   \bigg \}.
\end{eqnarray}

\subsection{The contribution $C_{2}$}

The contribution $C_2$ corresponds to the case where particle 2 is in the condensate, that is we calculate the collision integral with $f_2$ replaced by $(2\pi)^3 \delta^{(3)}(\p_2)$. We obtain
\begin{eqnarray}
C_2[n_c, g] 
 &=&   \frac{\kappa}{16\pi}\frac{n_c}{m} \,  \frac{1}{p_1 E_1}  \int_m^{E_1} dE_4 \, \frac{p_1^2 - (E_1+m)(E_4-m)}{[(E_4-m)+m_r]^2}  \nonumber \\ 
 && \qquad \times    {\bigg \{ } g_3 g_4 - g_1 g_3 -g_1 g_4 - g_1 {\bigg \} }{\bigg |}_{E_3 \to E_1+ m -E_4} 
\end{eqnarray}

The dominant contribution in the small scattering angle approximation corresponds now to the region $E_4\gtrsim m$. Accordingly, we set $\epsilon=E_4-m$ and $g_\epsilon \equiv g(E_4)$. Carrying out the expansion in powers of $\epsilon$ as we did earlier, we get
\begin{eqnarray}
C_2[n_c, g] 
 &=&  \frac{\kappa}{16\pi}\frac{n_c}{m} \,  \frac{1}{p_1 E_1}  \int_{\epsilon\to 0} d\epsilon \, \frac{p_1^2 - (E_1+m)\epsilon}{(\epsilon+m_r)^2}  \nonumber \\ 
 && \qquad \times    
 \left\{  	\left[  - (\epsilon g_\epsilon) g'_1  - g_1 (1+g_1) \right] +	\epsilon \, \left[ (\epsilon g_\epsilon)\frac{g''_1}{2}  +g_1 g'_1 \right]  \right\}
\end{eqnarray}
In a similar fashion as we treat $C_1$ above, we can perform the $\epsilon$-integrals and simplify the result to be: 
\beq
 {C}_2[n_c,g]  &=&\frac{\kappa}{16\pi}\frac{n_c}{m}   \bigg \{  \left(- K_a T^* - K_b \right ) \left( \frac{1}{m_r}\right) \nonumber \\
&& \quad +  
\left( -K_a V^* + K_c T^* + K_d \right )   \ln\left(\frac{\Lambda}{m_r}\right)   \bigg \}
\eeq
 with the $p_1$-dependent functions defined as 
\begin{eqnarray}\label{Kabcd}
&& K_a \equiv \frac{p_1 g?_1}{E_1},\qquad    K_b \equiv \frac{p_1 [g_1 (1+g_1)]}{E_1} \, , \,  \nonumber \\
 &&  K_c \equiv \frac{p_1 }{E_1} \left[\frac{g''_1}{2} + \frac{E_1+m}{p_1^2}g'_1\right],\quad     K_d  \equiv \frac{p_1}{E_1} \left[
  g_1 g'_1 +\frac{E_1+m}{p_1^2} g_1(1+g_1) \right]. \nn
\end{eqnarray}

\subsection{The contributions $C_{3,4}$}

Now we move to the cases with either $\p_3$ ($C_3$) or $\p_4$ ($C_4$) picking up the delta function component. As we have already noticed, the exchange of $\p_3$ and $\p_4$ corresponds to the exchange of the channels $u$ and $t$, with the symmetry $C_3^t=C_4^u$ and $C_3^u=C_4^t$. However, $C_3^t\ne C_3^u$, and the same holds for $C_4$.  This is because the  contribution to, say, $C_3^t$ is dominated by the region $\p_1\approx \p_3$, and $\p_3=0$.  As we shall see, a singularity develops then near $\p_1=0$. We shall then treat separately the two cases $C_3^t+C_4^u$ and $C_3^u+C_4^t$.

\subsubsection{$C_3^t+C_4^u$}
Let us first compute the $C_3^t$ and $C_4^u$ contributions following similar procedures as before: 
\begin{eqnarray}
C_3^t+C_4^u  &=&   \frac{\kappa}{16\pi}\frac{n_c}{m}\,   \frac{1}{p_1 E_1}  \int_m  dE_2 \, \frac{p_2^2 + (E_2-m)(E_4-E_2)}{[(E_1-m)+m_r]^2}  \nonumber \\ 
 && \qquad \times    {\bigg \{ }  - g_1 g_2 + g_1 g_4 + g_2 g_4 + g_4 {\bigg \} }{\bigg |}_{E_4 \to E_1+ E_2 - m} 
\end{eqnarray}
We   proceed to the small angle approximation and introduce the variable $\epsilon=E_1-m = E_4 - E_2$ and set $g_\epsilon \equiv g(E_1)$.  We obtain
\begin{eqnarray}
C_3^t+C_4^u 
 &=&    \frac{\kappa}{16\pi}\frac{n_c}{m}\,   \frac{1}{p_1 E_1} \int_{m} dE_2 \, \frac{p_2^2 + (E_2-m)\epsilon}{(\epsilon+m_r)^2}  \nonumber \\ 
 && \qquad \times    
 \left\{  	\left[    (\epsilon g_\epsilon) g'_2  + g_2 (1+g_2) \right] +	\epsilon \, \left[ (\epsilon g_\epsilon)\frac{g''_2}{2}  +  g'_2 (1+g_2) \right]  \right\} .\nn
\end{eqnarray}

As compared to the previous calculations, the singularity in the limit $\epsilon \to 0$ (or equivalently $p_1 \to 0$),  sits outside the energy integration. To handle this singularity, we perform an explicit integration over $\p_1$ in an infinitesimal sphere centered at $\p_1=0$. We get  
\begin{eqnarray}
\int_{p_1 \to 0} \{ C_3^t+C_4^u \}
 &=&    \frac{\kappa}{32\pi^3}\frac{n_c}{m}\int_{E_1 \to M} dE_1  \int_{M} dE_2 \, \frac{p_2^2 + (E_2-m)\epsilon}{(\epsilon+m_r)^2}  \nonumber \\ 
 && \quad \times    
 \left\{  	\left[    (\epsilon g_\epsilon) g'_2  + g_2 (1+g_2) \right] +	\epsilon \, \left[ (\epsilon g_\epsilon)\frac{g''_2}{2}  +  g'_2 (1+g_2) \right]  \right\} \nonumber \\ 
 &=&     \frac{\kappa}{32\pi^3}\frac{n_c}{m}   \left\{ H_a \,L_{g1} +  H_b \,L_0 + H_e  \,L_{g2}+ H_f \,L_1  \right\} \nonumber \\
&=&    \frac{\kappa}{32\pi^3}\frac{n_c}{m} \frac{1}{(2\pi^2)}  \bigg \{  \left(  H_a T^* +  H_b \right ) \left( \frac{1}{m_r}\right) \nonumber 
\\ && \qquad  +  
\left( H_a V^* + H_e T^* + H_f \right )   \ln\left(\frac{\Lambda}{m_r}\right)   \bigg \},
\end{eqnarray}
with 
\begin{eqnarray}
H_e & \equiv &  \int p^2 dE_p \,  \left[ \frac{g''(E_p)}{2} + \frac{(E_p-m)}{p^2} g'(E_p) \right] \\
H_f & \equiv &  \int p^2 dE_p \,  \left[ (1+g(E_p)) g'(E_p) + \frac{(E_p-m)}{p^2} g(E_p)\left (1+g(E_p) \right) \right] .
\end{eqnarray}
The result of this integration is finite. This allows us to identify a ``hidden'' singular  contribution to the condensate equation, of the form: 
\begin{eqnarray}
 \{ C_3^t+C_4^u \}  &\to&     (2\pi)^3 \delta^{3}(\p_1)      \frac{\kappa}{32\pi^3}\frac{n_c}{m} \bigg \{  \left(  H_a T^* +  H_b \right ) \left( \frac{1}{m_r}\right)    \nonumber  \\ 
 && \qquad \qquad +  \left( H_a V^* + H_e T^* + H_f \right )  \ln\left(\frac{\Lambda}{m_r}\right)   \bigg \}.
\end{eqnarray}

\subsubsection{$C_3^u+C_4^t$}
Let us now compute the $C_3^u$ and $C_4^t$ contribution: 
\begin{eqnarray}
C_3^u+C_4^t &=&    \frac{\kappa}{16\pi}\frac{n_c}{m}  \, \frac{1}{p_1 E_1}  \int_m  dE_2 \, \frac{p_1^2 + (E_1-m)(E_4-E_1)}{[(E_2-m)+m_r]^2}  \nonumber \\ 
 && \qquad \times    {\bigg \{ }  - g_1 g_2 + g_1 g_4 + g_2 g_4 + g_4 {\bigg \} }{\bigg |}_{E_4 \to E_1+ E_2 - m} .
  \end{eqnarray}
To proceed to the small angle approximation we introduce the variable $\epsilon=E_2-m = E_4 - E_1$ and set $g_\epsilon \equiv g(E_2)$. We then obtain: 
\begin{eqnarray}
C_3^u+C_4^t &=& \frac{\kappa}{16\pi}\frac{n_c}{m}  \, \frac{1}{p_1 E_1} \int_{\epsilon\to 0}  d\epsilon \, \frac{p_1^2 + (E_1-m)\epsilon}{(\epsilon+m_r)^2}  \nonumber \\ 
 && \quad 
 \left\{  	\left[    (\epsilon g_\epsilon) g'_1  + g_1 (1+g_1) \right] +	\epsilon \, \left[ (\epsilon g_\epsilon)\frac{g''_1}{2}  +  g'_1 (1+g_1) \right]  \right\} 
  \end{eqnarray}
After performing the $\epsilon$-integrals and we get 
\begin{eqnarray}
 C_3^u+C_4^t  &=&\frac{\kappa}{16\pi}\frac{n_c}{m}     \bigg \{  \left(  K_a T^* +  K_b \right ) \left( \frac{1}{m_r}\right) \nonumber \\
&& \quad +  
\left( K_a V^* + K_e T^* + K_f \right )   \ln\left(\frac{\Lambda}{m_r}\right)  \bigg \},
\end{eqnarray}
where the $p_1$-dependent functions are defined as 
\begin{eqnarray}
  K_e \equiv \frac{p_1 }{E_1} \left[\frac{g''_1}{2} + \frac{E_1-m}{p_1^2} g'_1\right], \;    K_f  \equiv \frac{p_1}{E_1} \left[
 (1+ g_1) g'_1 +\frac{E_1-m}{p_1^2} g_1(1+g_1) \right],
 \end{eqnarray}
and $K_a, K_b$ are given in Eq.~(\ref{Kabcd}).

\section{Cancellation of Leading Divergences and the Small Angle Equation}\label{Sect:leading log}

Finally let us put all pieces together. The results that we have obtained in the previous section exhibit divergences when one sends the screening mass to zero $m_r\to 0$. These divergence  are of two types: linear divergence $\sim 1/m_r$ and logarithmic divergence $\sim \ln[\Lambda/m_r]$. We shall see that all linear divergences cancel out.  This leaves only logarithmic singular terms,
leading to a consistent small angle approximation for both ``ordinary'' scattering involving regular 
distribution functions and for scattering to and from the condensate.

\subsection{Cancellation of Leading Divergences}
We first combine the contributions from $C_1= (2\pi)^3 \delta^{3}(\p_1) \tilde{C}_1$ and $C_3^t+C_4^u$, as follows: 
\begin{eqnarray}
C_1 + C_3^t+C_4^u &\to &    (2\pi)^3 \delta^{3}(\p_1)   \frac{\kappa}{32\pi^3}\frac{n_c}{m}  \bigg \{   \left[ \left(  -H_a T^* -  H_b \right ) +  \left(  H_a T^* +  H_b \right )\right] \left( \frac{1}{m_r}\right)    \nonumber  \\ 
 && \quad +  \left[ (-H_a + H_a ) V^* + (H_c+H_e) T^* + (H_d + H_f) \right ] \left[ \ln\left(\frac{\Lambda}{m_r}\right) \right] \bigg \}   \nonumber  \\
 &=&   (2\pi)^3 \delta^{3}(\p_1)  \frac{\kappa}{32\pi^3}\frac{n_c}{m}   \frac{1}{(2\pi^2)} \left[  H_1 T^* + H_2 \right]
  \left[ \ln\left(\frac{\Lambda}{m_r}\right) \right],
\end{eqnarray} 
with 
\begin{eqnarray}
H_1 & \equiv &  \int p^2 dE_p \,  \left[  g''(E_p)  + \frac{2E_p }{p^2} g'(E_p) \right] =   \int p^2 dp  \left\{  \frac{1}{p^2} \partial_p \left[ p^2 g' \right] \right\}   \\
H_2 & \equiv &  \int p^2 dE_p \,  \left[ (1+2g(E_p)) g'(E_p) + \frac{2E_p }{p^2} g(E_p)\left (1+g(E_p) \right) \right]  \nonumber \\
&=&   \int p^2 dp  \left\{  \frac{1}{p^2} \partial_p \left[ p^2  \, g(E_p)(1+g(E_p)) \right] \right\}.
\end{eqnarray} 

We then   combine the contributions from $C_2$ and $C_3^u+C_4^t$, as follows: 
\begin{eqnarray}
C_2 + C_3^u+C_4^t &\to & \frac{\kappa}{16\pi}\frac{n_c}{m} 
 \bigg \{ \left[  \left(- K_a T^* - K_b \right ) +  \left( K_a T^* + K_b \right )   \right ]\left( \frac{1}{m_r}\right) \nonumber \\
&& \quad +  
\left[ (-K_a + K_a ) V^* + (K_c +K_e) T^* + (K_d + K_f) \right ] \left[   \ln\left(\frac{\Lambda}{m_r}\right) \right] \bigg \} \nonumber \\
 &=&\frac{\kappa}{16\pi}\frac{n_c}{m}  \left[ K_1 T^* + K_2  \right]  \left[   \ln\left(\frac{\Lambda}{m_r}\right) \right],
\end{eqnarray}
with 
\begin{eqnarray}
  K_1 &\equiv& \frac{p_1 }{E_1} \left[ g''_1  + \frac{2E_1}{p_1^2} g'_1\right] = \frac{1}{p_1^2} \partial_{p_1} \left[ p_1^2 g'_1 \right]  \, , \\
       K_2  &\equiv& \frac{p_1}{E_1} \left[
 (1+ 2g_1) g'_1 +\frac{2E_1}{p_1^2} g_1(1+g_1) \right] 
 = \frac{1}{p_1^2} \partial_{p_1} \left[ p_1^2  \, g_1 (1+g_1) \right]. 
\end{eqnarray}

\subsection{Leading-log Equations}
Now we can put all pieces together, and obtain the following leading-log equations for the co-evolution of  the condensate and the regular distribution: 
\begin{eqnarray}
  \frac{\partial g(\p_1)}{\partial t} &=& \{-\nab {\cal J}_0 \}  
  +   \frac{ n_c}{m}    L_c \left\{   \frac{1}{p_1^2} \partial_{p_1}  \left[    p_1^2 (T^* g_1' + g_1(1+g_1)) \right]  \right \}   \label{eq_gp_final}   \nonumber \\ 
  \frac{\partial n_c}{\partial t} &=&   \frac{n_c}{m}  \frac{L_c}{(2\pi^2)}  \int p^2 dp  \left\{  \frac{1}{p^2} \partial_p  \left[    p^2 (T^* g' + g(1+g)) \right] \right\}  \nonumber \\
&& \quad    - \frac{1}{2\pi^2} \, \left[ p_0^2 S_0(p_0)\right]|_{p_0 \to 0}  \label{eq_nc_final} 
    \end{eqnarray}
    where $L_c$ is given in Eq.~(\ref{defLcmr}). 
The new terms associated with the condensate can be nicely rewritten after further introducing the following current: 
\beq\label{currentJc}
{\cal J}_c(\p_1) \equiv   - \,  \hat{\p}_1 \, L_c \,  \left[T^* g_1' + g_1(1+g_1) \right] \, \, .  
\eeq											
With the above, we can then rewrite the equations to be: 
\begin{eqnarray}
 \frac{\partial g(\p_1)}{\partial t} &=&  -\nab\cdot {\cal J}_0(\p_1) - \frac{n_c}{m} \nab \cdot {\cal J}_c(\p_1)  \label{eq_log_g}  ,  \\
  \frac{\partial n_c}{\partial t} &=&   - \frac{1}{2\pi^2} \, \left[ p_0^2 {\cal J}_0(p_0)\right]_{p_0 \to 0}  - \frac{1}{2\pi^2} \, \frac{n_c}{m} \, \left[ p_0^2 {\cal J}_c(p_0)\right]_{p_0 \to 0} \label{eq_log_c},
\end{eqnarray}
where it is understood that the first equation is valid for all $p_1>0$.

Lastly let us analyze behavior of the Eqs.(\ref{eq_log_g},\ref{eq_log_c}) at extremely small momentum regime. We focus on the evolution after the onset of condensation, i.e. with a nonzero condensate $n_c>0$. As already analyzed previously, the behavior of $g(p)$ at very small p is a local thermal form plus certain corrections, i.e. 
\begin{eqnarray}
g(p\to 0)= g^*(p)+ \delta g(p)    ,\qquad   g^*(p)\equiv  \frac{1}{e^{(E-M)/T^*}-1}      \label{eq_g_local}
\end{eqnarray}
with $T^*=I_a/I_b$ and $\delta g(p)$ is supposed to be regular as $p\to 0$, with $\delta g(0)$ finite. 

Let us recall that before onset, $n_c=0$, and ${\cal J}_0\sim p$ at small $p$ so that only the first equation contributes. Just before the onset the current ${\cal J}_0$ contains, besides the term $\sim p$ at very small $p$ singular contributions in $1/p$ and $1/p^2$. The singularity in $1/p^2$ dominates, but disappears just at onset (when $\mu=m$), leaving a contribution ${\cal J}_0\sim 1/p$. It follows that, after onset,  $\left[ p_0^2 {\cal J}_0(p_0)\right]_{p_0 \to 0}=0$, and only ${\cal J}_c$ contributes in the equation for $n_c$, i.e., Eq.~(\ref{eq_log_c}) reads
\begin{eqnarray}\label{eq_log_c2}
  \frac{\partial n_c}{\partial t} =   - \frac{1}{2\pi^2} \, \frac{n_c}{m} \, \left[ p_0^2 {\cal J}_c(p_0)\right]_{p_0 \to 0} \label{eq_log_c}.
\end{eqnarray}
Since $T^*=I_a/I_b$, the contribution to ${\cal J}_c$ in Eq.~(\ref{currentJc})  comes from  terms of the form $\delta g_1\, g^*_1\sim 1/p^2$ (assuming $\delta g_1(0)$ is finite), so that $
{\cal J}_c(p\to 0) \sim \hat\p_1 /{p^2}$, which yields indeed a finite contribution to Eq.~(\ref{eq_log_c}). Note also that while the surface term $[p_0^2 J_0(p_0)]_{p_0->0}$ scales as $~p_0 $ and hence does not contribute to the condensate evolution, it does control the evolution of $g(p=0)$ in Eq.~(\ref{eq_log_g}), since the integral of $g(p)$ in a small sphere of radius $p_0$ also scales as $p_0$.

As a consistency check, one can easily verify that the two equations (\ref{eq_log_g})  and (\ref{eq_log_c2}) satisfy the conservation laws of particle number and energy.  Consider first particle number conservation: 
\begin{eqnarray}
\partial_t \left\{ \int_{\p} g(\p)  + n_c \right\} = 0 .
\end{eqnarray} 
Because the currents vanish when $\p\to\infty$, the result of integrating the right hand side of Eq.~(\ref{eq_log_g}) is  
\beq
\frac{1}{2\pi^2} \, \left[ p_0^2 {\cal J}_0(p_0)\right]_{p_0 \to 0}  + \frac{1}{2\pi^2} \frac{n_c}{m}  \left[ p_0^2 {\cal J}_c(p_0)\right]_{p_0 \to 0}. 
\eeq
The first term vanishes, as we have argued above. The second term  cancels with the right hand side of Eq.~(\ref{eq_log_c2}).
For the energy conservation, we need to examine the following: 
\begin{eqnarray}\label{energyconserv1}
&&\partial_t \left\{ \int_{\p}  E_\p\,  g(\p)  + m n_c \right\} =  \int_{\p} E_\p\, \left[-\nab\cdot {\cal J}_0(\p) \right] +   \frac{n_c}{m} \int_{\p} E_\p\, \left[ -\nab\cdot {\cal J}_c(\p)  \right]  \nonumber \\
&& \qquad\qquad\qquad\qquad\qquad\qquad  - \frac{n_c}{2\pi^2} \,  \left[ p_0^2 {\cal J}_c(p_0)\right]_{p_0 \to 0} .
\end{eqnarray} 
The first term on the right hand side may be written as
\begin{eqnarray}
\int_{\p} E_\p\, \left[-\nab\cdot {\cal J}_0(\p) \right] &=& \int_{\p} \left\{ -\nab\cdot \left[E_\p {\cal J}_0(\p) \right]  + \frac{\p}{E_\p}\cdot {\cal J}_0(\p) \right \}  \nonumber \\ 
&=& \frac{m}{2\pi^2} \, \left[ p_0^2 {\cal J}_0(p_0)\right]_{p_0 \to 0} + \int_{\p}  \frac{\p}{E_\p}\cdot {\cal J}_0(\p)\nn
&=&0,
\end{eqnarray}
where we have used  $\left[ p_0^2 {\cal J}_0(p_0)\right]_{p_0 \to 0}=0$, and the proof that the second term vanishes is identical to that of energy conservation before onset (see paper 1). 
For the second term of Eq.~(\ref{energyconserv1}), we have
\begin{eqnarray}
 \frac{n_c}{m} \int_{\p} E_\p\, \left[ -\nab\cdot {\cal J}_c(\p)  \right] &=&  \frac{n_c}{m} \int_{\p} \left\{ -\nab\cdot \left[ E_\p {\cal J}_c(\p)  \right] + \frac{\p}{E_\p}\cdot {\cal J}_c(\p) \right\}  \nonumber \\ 
 &=& \frac{m}{2\pi^2} \, \frac{n_c}{m} \, \left[ p_0^2 {\cal J}_c(p_0)\right]_{p_0 \to 0} + \frac{n_c}{m} \,   \int_{\p}  \frac{\p}{E_\p} \cdot{\cal J}_c(\p) \nn
 &=&\frac{n_c}{2\pi^2} \, \left[ p_0^2 {\cal J}_c(p_0)\right]_{p_0 \to 0} , 
\end{eqnarray}
where, in the last step,  we have used the explicit expression (\ref{currentJc}) of the current, with $T^*=I_a/I_b$ and $I_a$, $I_b$ defined in Eq.~(\ref{IaIbdef}). This result is equal and opposite to the last term of Eq.~(\ref{energyconserv1}), which completes the proof of energy conservation.

\section{Conclusion}

The results of our analysis are  Eqs.~(\ref{eq_log_g}) and (\ref{eq_log_c2}), or equivalently Eq.~(\ref{transportfinal}).  We have shown that these equations 
are of leading order in the small angle approximation (leading logarithm of the Debye screening mass). These equations satisfy energy and momentum conservations,  and have a thermal fixed point that consists of a finite condensate plus a Bose-Einstein distribution with critical chemical potential $\mu=m$. The behavior of these equations in the vicinity of zero momentum was carefully analyzed, exploiting results reported in a companion paper \cite{BJL}, and the main singular behavior of the particle current near the origin in momentum space was  identified. 
Numerical analysis of these equations will be reported in a later paper.

\section*{Acknowledgements}
 We would like to express our warmest thanks to Larry McLerran for numerous discussions and insightful remarks. The research of JPB is supported by the European Research Council under the Advanced Investigator Grant ERC-AD-267258. That of JL is supported by the National Science Foundation under Grant No. PHY-1352368. JL is also
grateful to the RIKEN BNL Research Center for partial support.

\end{document}